\documentclass[a4,aps,twocolumn,showpacs,preprintnumbers,amsmath,amssymb]{revtex4} 
\usepackage{dcolumn}
\usepackage{graphicx}
\usepackage{amsfonts} 
\usepackage{epstopdf}
\usepackage{color}
\usepackage{here}

\begin{document} 
\title{Vortex excitation in a stirred toroidal Bose-Einstein condensate}
\date{\today}
\author{A.I. Yakimenko$^{1}$, K.O. Isaieva,$^1$, S.I. Vilchinskii$^1$, E.A. Ostrovskaya$^{2}$}
\affiliation{$^1$Department of Physics, Taras Shevchenko National University of Kyiv, 64/13, Volodymyrska Str. City of Kyiv, 01601, Ukraine \\
$^2$Nonlinear Physics Centre, Research School of Physics and Engineering, The Australian National University, Canberra ACT
0200, Australia}

\begin{abstract}
Motivated by the recent experiment [Wright {\em et al.}, Phys. Rev. A {\bf 88}, 063633 (2013)], we investigate formation of vortices in an annular BEC stirred by a narrow blue-detuned optical beam. In the framework of a two-dimensional mean field model, we study the dissipative dynamics of the condensate with parameters matched to the experimental conditions. Vortex-antivortex pairs appear near the center of the stirrer  in the bulk of the condensate for slow motion of the stirring beam. When the barrier angular velocity is above some critical value, an outer edge  surface mode develops and breaks into the vortices entering the condensate annulus.
We determine the conditions for creation of the vortex excitations in the stirred toroidal condensate and compare our results with the experimental observations.
\end{abstract}

 \maketitle

\section{Introduction}
Nucleation of quantized vortices and their dynamics are closely associated with supeflow decay and turbulence in quantum fluids and atomic Bose-Einstein condensates (BECs) \cite{FetterReview2001}.
In early experiments, a simple-connected geometry of the trap was used to measure onset of dissipationless flow in BEC. A  laser beam was moved through the condensate, and the dissipation was detected as heating of the condensate \cite{PhysRevLett.83.2502,PhysRevLett.84.806,PhysRevLett.85.2228,PhysRevLett.87.210402,PhysRevLett.104.160401}. The problem of the vortex excitation by a moving barrier was also extensively investigated theoretically \cite{PhysRevLett.80.3903,PhysRevA.61.051603,PhysRevA.62.061601,PhysRevX.1.021003}.

Recent experimental progress in creating atomic gases in a toroidal geometry has opened novel prospects for the studies of the fundamental properties of the superfluid state \cite{PhysRevA.88.053615}.
Ring-shaped BEC has become the topic of a large body of experimental and theoretical research
\cite{PhysRevA.66.053606,Benakli1999,Brand01,PhysRevA.74.061601,Das2012, PhysRevA.64.063602,Dalfovo06}
including  persistent currents \cite{PhysRevLett.99.260401,PhysRevLett.110.025301,PRA2013R}, weak links \cite{PhysRevLett.106.130401,PhysRevLett.110.025302,PhysRevA.80.021601,Piazza2013}, solitary waves \cite{Brand01,Berloff09}, and decay of the persistent current via phase slips \cite{PhysRevA.86.013629}. A persistent flow in a toroidal trap can be created  by transferring a quantized angular
momentum from optical fields \cite{PhysRevLett.99.260401,PhysRevLett.110.025302} or by stirring with rotating blue-detuned laser beam \cite{PhysRevLett.110.025302,Wright2013}. The quantized circulation in a ring
corresponds to a $m$-charged vortex line pinned at the center of the ring-shaped condensate, where the vortex energy has a local
minimum.
Since the vortex core is bounded by the potential barrier, even the multicharged ($m>1$)
metastable vortex states can be very robust in a ring-shaped condensate \cite{PhysRevA.86.013629}.

Very recently \cite{Wright2013}, vortices in a toroidal trap were excited using a  \emph{small} (diameter less than the width of the annulus) variable-height potential barrier (a 'stirrer') with an angular velocity ranging from zero up to the speed of sound in the condensate. A wide range of the experimental parameters used in Ref. \cite{Wright2013} opens an intriguing possibility for theoretical investigation of excitation of the persistent current in a stirred ring-shaped BEC at different regimes. A simple 1D model used in Ref. \cite{Wright2013} shows a reasonable agreement with the experimental measurements of the threshold for vortex excitation. However, the microscopic origin of the superflow generation, as well as complex vortex dynamics, are beyond the scope of the 1D treatment, which is based on the analysis of the average speed of sound. In this work, we consider a 2D reduction of the full 3D mean-field model justified by the trapping geometry. Performing numerical modeling for various combinations of potential barrier height and angular velocity, we determine the conditions for creation of vortices and compare it with the experimental findings.

\section{Model}\label{model}
Here we formulate the dissipative mean-field model used in our work.
In modeling nonequilibrium behavior, such as nucleation
of vortices, dissipative effects are of crucial importance
since they provide the mechanism for damping of elementary
excitations in the process of relaxation to an equilibrium state.
It is the dissipation that either causes the vortex line to drift
to the outer edge of the condensate (where vortices decay) or leads
to the pinning of the vortex in the central hole of the ring-shaped condensate. The relaxation of the vortex core position to the
local minimum of the energy leads to formation of the metastable persistent current.
Dissipative effects appear in a
trapped condensate due to interaction with thermal cloud
and can be captured phenomenologically by the dissipative Gross-Pitaevskii  equation (GPE) \cite{Pitaevskii59,Choi98,Tsubota13}.
For a system of weakly interacting degenerate atoms close to
thermodynamic equilibrium and subject to weak dissipation,
the dissipative GPE for the macroscopic wavefunction can be written in the form
\begin{equation}\label{GPE}
(i-\gamma) \hbar \frac{\partial \tilde\Psi(\textbf{r},t)}{\partial t} = \left[\hat H + U_0 |\tilde\Psi(\textbf{r},t)|^2 -\mu\right]\tilde\Psi(\textbf{r},t),
\end{equation}
where $\gamma\ll 1$ is the phenomenological dissipative parameter, $\hat H=-\frac{\hbar^2}{2M} \Delta + V(\textbf{r})$,  $\Delta$
is a Laplacian operator, $U_0 = \frac{4 \pi \hbar^2 a_s}{M}$ is interaction strength, $M$ is the mass of the $^{23}$Na atom, $a_s=2.75$nm is the $s$-wave scattering length. The chemical potential $\mu(t)$ of the equilibrium state in our dynamical simulations was adjusted at each time step so that the number of condensed particles slowly decays with time: $N(t)=N(0)e^{-t/t_0}$, where $t_0 = 10$ s corresponds to the $1/e$ lifetime of the BEC reported in Ref. \cite{PhysRevLett.110.025301}.

The trapping potential
$$V(\textbf{r})=V_{\textrm{tr}}(r,z)+V_{\textrm{b}}(x,y,t),$$
consists of an axially-symmetric toroidal trap:
$$V_{\textrm{tr}}(r,z)=\frac12 M\omega_z^2 z^2+\frac12 M\omega_r^2(r-R)^2,$$
where $r=\sqrt{x^2+y^2}$,
and the repulsive potential of optical blue-detuned stirring beam:
$$V_{\textrm{b}}(x,y,t)=f(t)e^{-\frac{1}{2 d^2}\left\{[x-x_0(t)]^2+[y-y_0(t)]^2\right\}},$$
 where $\left\{x_0,y_0\right\}=\left\{R\cos(\Omega t),R\sin(\Omega t)\right\}$ is the coordinate of the barrier center, moving counter-clockwise with a constant angular velocity $\Omega$ through the maximum of condensate density [see Fig. \ref{GroundState} (a)]. The effective width of the beam $d$ is significantly less than the width of the annulus $\Delta R$, but is ten times greater than the healing length $\xi$ ($d=3.82\mu$m, $\Delta R> d\gg\xi=1/\sqrt{8\pi n_0 a_s}\approx 0.32\mu$m, where $n_0\approx 1.4\cdot 10^{14}$ cm$^{-3}$ is condensate peak density).  The function $f(t)$ describes the temporal  changes in the barrier height: $f(t)$ linearly ramps up  in the first $0.1$ s of evolution from zero to $U_b$ and remains unchanged for $0.8$ s, then in the last $0.1$ s it ramps down to zero again.
The radius of the ring trap is equal to $R = 22.6\mu$m.
The experimentally measured trapping frequencies are $\omega_r=2\pi\times 134$ Hz,
  $\omega_z=2\pi\times 550$Hz, which corresponds to the oscillator lengths $l_r=\sqrt{\hbar/(M\omega_r)}= 1.81\mu$m and $l_z=\sqrt{\hbar/(M\omega_z)}= 0.90\mu$m. 

The harmonic potential creates a tight binding confinement in $z$ direction, so that the BEC cloud is 'disk-shaped'.
Consequently, we assume that the longitudinal motion of condensate is frozen, and the wavefunction can be factorized as follows $\Psi_j(\mathbf{r},t)=\tilde\Psi_j(x,y,t)\Upsilon(z,t),$ where
$\Upsilon(z,t)=(l_{z}\sqrt{\pi})^{-1/2}\exp(-\frac{i}{2}\omega_zt-\frac12z^2/l_{z}^2)$. The norm of the condensate wave  function $\tilde\Psi(x,y)$ is equivalent to the number of atoms:
$
N=\int|\tilde\Psi|^2dx dy.
$
After integrating out the longitudinal coordinates in the 3D GPE (\ref{GPE}),  we obtain a dissipative GPE in 2D:
\begin{equation}\label{GPE_dissipative1}
(i-\gamma)\frac{\partial \psi}{\partial t} = \left[-\frac{1}{2} \Delta_\perp + V(x,y,t)+  g|\psi|^2 -\mu\right]\psi,
\end{equation}
where $V(x,y,t)=\frac12(r-R)^2+V_\textrm{b}(x,y,t)$ and  $R=12.47$ are the dimensionless potential and the radius of the trap,
$g=\sqrt{8\pi}a_s/l_z=1.54\times 10^{-2}$ is the dimensionless 2D interaction constant, $\psi\rightarrow l_r \psi$ is dimensionless wave-function.
Here we use  harmonic oscillator units: $t\to \omega_r t$, $(x,y)\to (x/l_r,y/l_r)$, $V_\textrm{b}\to V_\textrm{b}/(\hbar\omega_r)$, $\mu\to \mu/(\hbar\omega_r)$.

 \section{Ground state: stationary solutions and excitation spectrum}\label{SecGroundState}
\begin{figure}[h]
  \includegraphics[width=3.4in]{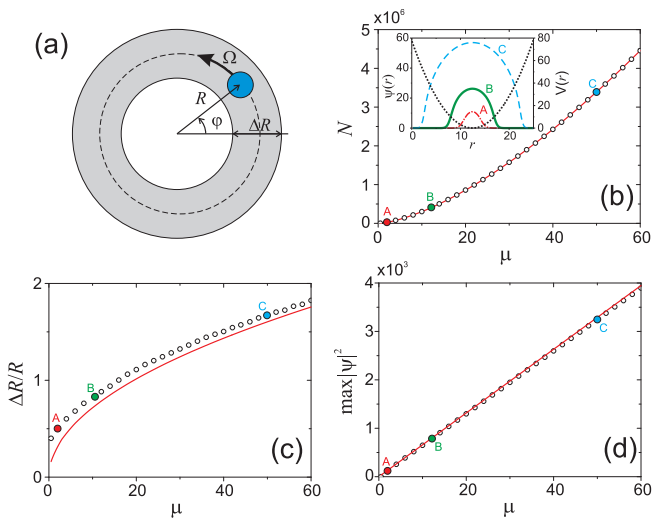}
  \caption{(Color online) (a) Scheme of the 2D BEC
  in a ring trap  of radius $R$. A blue-detuned optical stirrer (blue filled circle) excites vortices when it circles   counter-clockwise with angular velocity $\Omega$ through the maximum of condensate density (black dashed line). Stationary ground states: (b) Number of atoms vs chemical potential. Inset shows the radial trapping potential $V(r)$ (black dotted line) and examples of numerically found radial profiles $\psi(r)$ for three values of $\mu$. A: $\mu=2$, B: $\mu=10.62$, and C: $\mu=50$. (c) Solid red lines shows  relative Thomas-Fermi width $\Delta R/R$ of the ring  as a function of chemical potential. Circles represent numerical results for the relative width $\Delta R/R$ of the condensate with density above $10^{-3}$ of the peak density. (d) Peak density max$|\psi|^2$ vs $\mu$. Solid red line presents the results of TF approximation, open circles correspond to the numerical stationary solutions.}
  \label{GroundState}
\end{figure}

The radially-symmetric steady-states  in toroidal BEC, corresponding to vortices with a topological charge $m$ were found numerically integrating a conservative stationary GPE
$$\mu\psi_m(r)=-\frac12\Delta_r^{(m)}\psi_m(r)+\frac12(r-R)^2\psi_m(r)+g\psi_m^3(r),$$ where
$$\Delta_r^{(m)}=\frac{d^2}{d r^2}+\frac{1}{r}\frac{d}{dr}-\frac{m^2}{r^2}.$$ A ground state with $m=0$ corresponds to boundary conditions $\psi'(0)=0$, $\psi(\infty)=0$. The numerical results for stationary ground states are given in Fig. \ref{GroundState}. Using the Thomas-Fermi (TF) approximation, it is straightforward to obtain the approximate expression for the number of atoms:
$N=\frac{4R}{3a_s}\left(\pi\mu^3\omega_r/\omega_z\right)^{1/2},$ width of the BEC annulus: $\Delta R= 2\sqrt{2\mu}$, and the peak density max$|\psi|^2=\mu/g$. These simple estimates are found to be in excellent agreement with numerical results (see Fig. \ref{GroundState}). As can be seen, the radial profiles gradually expand when chemical potential grows and for $\mu>60$ the central hole is filled by the atoms, so that the topology of the condensate becomes single-connected.

Let us discuss the choice of parameters of our theoretical model. As pointed out above, the BEC cloud is highly anisotropic in the experimental setup, thus the 2D model properly describes the main features of the vortex dynamics. However, it is easy to verify that, if one uses the  harmonic oscillator ground state in $z$-direction,
 then  the peak density $n_0$ in $z=0$ plane becomes overestimated compared to the actual experimental value for the same value of total number of atoms. Obviously, in the inhomogeneous condensate, excitation and evolution of the vortices are sensitive rather to the condensate density structure than to the total number of atoms. Parameters of our 2D model are matched to the experimental value of the peak density and geometry of the experimental setup.
Corresponding dimensionless value for the chemical potential is $\mu=10.62$. The stationary solution for the parameters of the model used in our simulations is shown in the insight in Fig. \ref{GroundState} (a) and in Fig. \ref{GroundState} (b)-(d) by the green circle. Note that, for the parameters under consideration, the central hole of the ring-shaped condensate is well pronounced, and the width of the annulus $\Delta R=9.22$ is much greater then the effective width of the barrier $d=2.1$.

\begin{figure}[h]
  \includegraphics[width=3.4in]{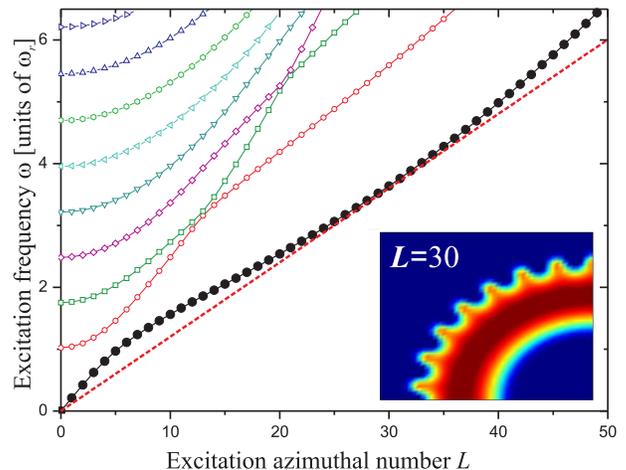}
  \caption{(Color online) Excitation spectrum $\omega(L)$ of the ground state ($m=0$) of toroidal condensate. Dashed straight line corresponds to $\omega=L\Omega_c$, where the critical angular velocity $\Omega_c$ is obtained in the frame of the surface model. Note that the critical angular velocity $\Omega_c=$ min $\left\{\omega(L)/L\right\}$ obtained from the Bogoliubov spectrum predicts the same value: $\Omega_c\approx 2\pi\times 16.1$Hz. Inset shows the fragment of the annulus with perturbed atomic density for the critical outer edge surface mode ($L=30$, size of the image is $20\times 20$ in units of $l_r$). The amplitude of the perturbation is normalized to be 2$\%$ of the peak density of the unperturbed condensate.}
  \label{Spectrum}
\end{figure}

The analysis of the elementary excitations of the ground state provides an insight into stability properties of the stirred condensate. In the toroidal geometry the radial degrees of freedom lead to specific features of the excitations, such as inner and outer edge surface modes, which cannot be explained in terms of the average speed of sound \cite{PhysRevA.86.011604,PhysRevA.86.011602}.
As noted in Ref. \cite{stuartthesis}, if the geometry of the toroidal trap is fixed, then the nature of the excitation is determined by the number of atoms in the condensate. For a large condensate an outer edge surface mode is dominating, while  for a weak condensate the instability changes to a phonon mode, located in the condensate bulk. To verify which type  of the excitations dominates in our setup, and to obtain the quantitative characteristics of the corresponding eigenfrequencies, we have performed the analysis of Bogoliubov azimuthal spectrum for parameters of the condensate matched to the experimental conditions \cite{Wright2013}.
Expanding the nonstationary solution in the vicinity of a $m$-charged steady state $\psi(x,y,t)=\left[\psi_m(r)+\delta\psi(r,\varphi,t)\right]e^{-i m\varphi}$, where $\delta\psi=u(r)e^{i\omega t + iL\varphi}+v^*(r)e^{-i\omega^* t- iL\varphi}$  and $L$ is the integer azimuthal mode number, and linearizing the dynamical Eq. (\ref{GPE_dissipative1}) without dissipation ($\gamma=0$ and $\mu=$const) we obtain the eigenvalue problem for $\omega$ of the form:
 \begin{equation}
 \label{eq:EigenProblem}
  \begin{pmatrix}
    \hat Q_{m+L} & -g\psi_m^2 \\
g\psi_m^{*2}     & -\hat Q_{m-L}
  \end{pmatrix}
\vec{\varepsilon}=\omega\vec{\varepsilon},
 \end{equation}
where $\vec\varepsilon=(u,v)$,
$$\hat Q_{m\pm L}=\mu+
\frac12\Delta_r^{(m\pm L)}-\frac12(r-R)^2-2g|\psi_m(r)|^2.$$  The unperturbed radial profile $\psi_m(r)$ can be considered as a
real function without loss of generality. Here we follow the descritization procedure described in Ref.\cite{stuartthesis} in order to reduce the differential eigenvalue problem (\ref{eq:EigenProblem}) to the linear algebraic one. Figure \ref{Spectrum} presents the results our calculations of the Bogoliubov spectrum that are similar to the ones reported in Refs. \cite{PhysRevA.86.011604,PhysRevA.86.011602}. Note that all frequencies are real-valued.

It appears that the linear eigen-modes corresponding to the critical angular frequency $\Omega_c=\textrm{min}\left\{\omega(L)/L\right\}$, obtained by analogy with the Landau criterion (see, e.g., Ref. \cite{DalfovoStringari2000}), are localized at the external surface of the annulus. The corresponding perturbed density distribution is shown in the inset of the Fig. \ref{Spectrum}, where perturbation $\delta n=2\textrm{Re}\left[\psi_m^*\delta\psi\right]$ is normalized so that $\delta n/n_0\le 2\cdot 10^{-2}$, where $n_0$ is the condensate peak density. It is remarkable that the critical frequency $\Omega_c/\omega_r=\sqrt{2}\mu^{1/6}/(R+\Delta R/2)$ obtained in the frame of the simple surface model (see, e.g., \cite{PhysRevA.86.011602}), where the curvature of the external surface is neglected, gives practically the same value $\Omega_c=2\pi\times 16.1$Hz.

Since the stirring by the laser beam is accompanied by the condensate distortion and shape deformation, one could expect that a sufficiently rapid barrier (with $\Omega>\Omega_c$)  produces an excited state corresponding to a surface wave with the energy $\hbar\omega(L)$ and angular momentum $\hbar L$ along $z$. Once these surface waves are excited, they break up and nucleate vortices, which come into the condensate bulk from the outer boundary of the annulus. For a slow barrier (with $\Omega<\Omega_c$) the outer edge surface mode is not excited and the dominating mechanism for the vortex nucleation is expected to be the formation of vortex pairs in the condensate bulk. The direct numerical simulations of the dissipative dynamics of the stirred condensate support these predictions of the linear stability analysis.

\section{Dissipative dynamics of stirred toroidal condensate}
\begin{figure}[h]
  \includegraphics[width=3.4in]{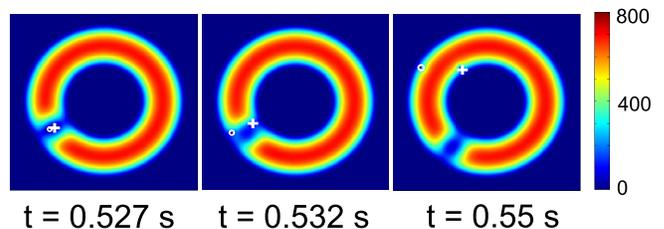}
  \caption{(Color online) Excitation of a vortex-antivortex pair by the stirring laser beam with  $\Omega =2\pi\times 3$ Hz and $U_b/h=1300$ Hz. The core of the vortex is indicated by a cross; antivortex is marked by a circle.}
  \label{Dynamics3Hz}
\end{figure}

\begin{figure}[h]
  \includegraphics[width=3.4in]{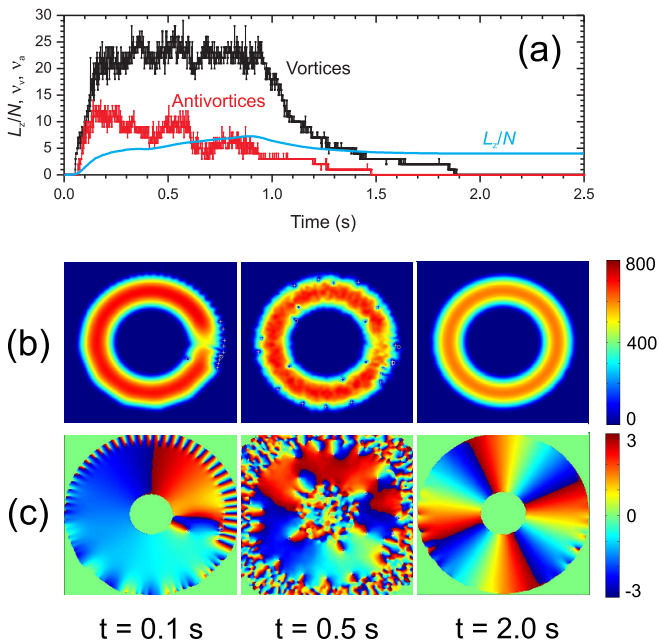}
  \caption{(Color online) Temporal evolution of the toroidal condensate with fast stirrer ($\Omega =2\pi\times 20$ Hz, $U_b/h=300$ Hz): (a) The angular momentum per atom $L_z/N$ (blue curve), number of annular vortices $\nu_v$ (black dots), and  antivortices $\nu_a$ (red dots). (a), (b) Snapshots of the density and phase distribution in $(x,y)$ plane for different moments of time (the size of each image is $40\times 40$ in units of $l_r$).  }
  \label{Dynamics20Hz}
\end{figure}

Here we present the results of our numerical simulations of the dissipative GPE (\ref{GPE_dissipative1})
with the split-step Fourier transform method.  In line with the experimental procedure in our calculations, initially nonrotating condensate is stirred for one second with a repulsive potential moving azimuthally at a fixed angular velocity.  In the following, we neglect possible position and temperature dependence of
$\gamma$ and set $\gamma = 1.5\cdot 10^{-3}$. We note that our main results do not depend
qualitatively on the chosen value of $\gamma\ll 1$. The dissipative parameter $\gamma$ determines the relaxation time of the vortices: the greater $\gamma$ is, the less time it takes for a vortex to drift from the high-density condensate annulus to the low-density periphery. With $\gamma = 1.5\cdot 10^{-3}$, we found the bulk of the condensate annulus to be cleansed from the vortices after three seconds, which is consistent with the experimentally measured lifetime of the annular vortices \cite{Wright2013}. Typical examples of the dissipative dynamics are given  in Fig. \ref{Dynamics3Hz} for slow and in Fig. \ref{Dynamics20Hz} for fast rotation of the barrier.
More examples of evolution of the density and phase of the toroidal condensate for different angular velocities and heights
of the stirring beam are given in the Supplemental Material.

We observed two regimes of the vortex excitation in a toroidal BEC.  For a {\em low angular velocity} of the stirrer vortex-antivortex pair are nucleated near the center of the barrier (see Fig. \ref{Dynamics3Hz}). Then the pair undergoes a breakdown and the antivortex moves spirally to the external surface of the condensate and finally decays into elementary excitations, while the vortex becomes pinned in the central hole of the annulus adding  one unit to the topological charge of the persistent current. The progressive drift of the vortices toward the external condensate boundary is the result of dissipation, which leads to the vortex energy decay. 

It is interesting to note a considerable  phase gradient traveling in front of the rotating barrier even for  a low barrier amplitude and slow rotation,  long before the vortex pair appears (see the animations from Supplemental Material). The phase gradient corresponds to the velocity field, which means that a  forward superfluid flow appears in the stirred condensate even well below the threshold rotation rate. This specific feature of the multiply-connected superfluid has been conclusively established experimentally in Ref. \cite{arxiv1406_1095} and can be easily explained as follows: a low-density wake appearing directly behind the stirrer initiates two counter-propagating superflows. First flow is directed against the rotation of the barrier. This atomic flow tends to fill the low-density wake behind the barrier. Because the condensate wave-function must be single valued, the velocity circulation around any closed path must be a multiple of $2\pi m$, thus the total velocity circulation vanishes for the state with zero winding number ($m=0$). The second atomic flow cancels the phase gradient corresponding to the velocity field of the atoms moving through the barrier. That is why, in a ring-shaped BEC, a  superflow, which is co-directional with the stirrer, appears  even for low rotation rates.

It turns out that, if the barrier intensity $U_b$ is just above the threshold value, then the angular momentum transfers to the condensate only by nucleation of the vortex-antivortex pair in the bulk of the annulus. However, if $U_b$ is much greater than the threshold value, then a weak link (a localized region of rarefied superfluid density) develops in a ring and breaks the potential barrier for the external vortices. As a result, a vortex from the outside of the ring can enter though the rotating weak link. This mechanism is similar to stirring with a wide barrier observed in Ref. \cite{PhysRevLett.110.025302}.


For {\em higher angular velocities} of the stirrer (with $\Omega>\Omega_c$, where $\Omega_c$ is found to be close to the one predicted by Bogoliubov analysis) the dominating source of vortices is the instability of the external surface modes.  First, ripples appear at the external surface  (see Fig. \ref{Dynamics20Hz}), and then several vortices nucleate simultaneously. Also, as often happens for a higher barrier intensity $U_b$, vortex lines come into the bulk of the condensate through the rotating weak link. Further complex dynamics of the vortices is governed not only by the condensate inhomogeneity and dissipation effects, but also by the interplay between
condensate flows corresponding to other vortices.
Vortex-antivortex annihilation produces sound waves, which also interact with other vortices. Moreover, the sound pulses can also break into vortex-antivortex pairs. As the barrier circles in the annulus, it periodically travels through its own low-density wake, creating new vortices and interacting with the existing ones. The number of vortices increases dramatically in our simulations with increasing barrier amplitude $U_b$, so that dynamics of the vortices becomes quite irregular. Note that in the central hole and far from the external surface of the condensate, where the condensate density vanishes, the phase strongly fluctuates and one can see infinitely many vortices and antivortices. To avoid the influence of these ``ghost vortices'', we discard the contribution  to $\nu_v$ and $\nu_a$ of the vortices and antivortices which are located beyond the TF surface.  Since a host of vortices is located at the periphery of the condensate, the number of annular vortices $\nu_v$ and antivortices $\nu_v$ fluctuates when they cross the boundary of the TF surface [see Fig. \ref{Dynamics20Hz} (a)]. To illustrate the relaxation process and formation of a persistent current in Fig. \ref{Dynamics20Hz}, the time-frame of the numerical modeling was substantially extended in comparison with the experiment.

 \begin{figure}[h]
  \includegraphics[width=3.4in]{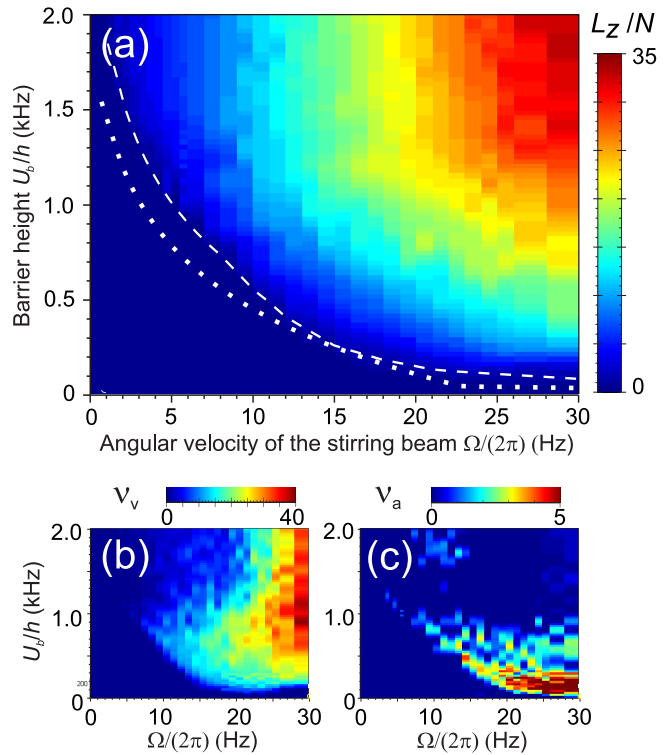}
  \caption{(Color online) Angular momentum, number of annular vortices and antivortices for the final stage of the stirring process ($t=1$ s) are represented by the color code as a function of angular speed $\Omega$ and the height of the stirring potential $U_b$: (a) Angular momentum per atom $L_z/N$. White isolines correspond to $L_z/N=1$ (dashed curve) and $L_z/N=0.02$ (dotted curves). (b) Number of annular vortices $\nu_v$. (c) Number of annular antivortices $\nu_a$.}
  \label{UbOmega}
\end{figure}

Vortex excitations experience dissipation all the time, which sends them in search of  an equilibrium position in accordance with  the local minima of the vorticity energy.
In the inhomogeneous ring-shaped condensate, the two options are either relaxation to the persistent current, when vortex line is pinned in the giant central hole, or to the outer boundary, where vortices decay into elementary excitations.
As is well known (see, e.g. \cite{Tsubota13}), small-scale forcing may generate large-scale flows in effectively 2D turbulent classical and quantum fluids. This feature is in a sharp contrast to hydrodynamic turbulence in 3D fluids. Recently \cite{PhysRevLett.111.235301} it was demonstrated, both experimentally and numerically, for an oblate Bose-Einstein condensate confined in an annular trapping potential, that inverse energy cascades are inherent for 2D quantum turbulence.
 In this context formation of large-scale persistent current  observed in our simulations is naturally determined by the development of the 2D turbulence. The energy of small-scale forcing transits first to irregular vortex distributions and then relaxes to a large-scale flow in the form of a circulating superflow. For example, as is seen from Fig. \ref{UbOmega} (a), the number of vortices and antivortices rapidly increases when the stirring beam ramps up ($0<t<0.1$ s), and decays to zero when the stirring beam is switched off ($t>1$ s), while the angular momentum per atom saturates to an integer number $m=4$ corresponding to the $m$-charged persistent current.  

We have performed an extensive series of numerical simulations of a full one second operating cycle for different values of barrier intensity $U_b$ and angular velocity $\Omega$. To summarize our findings and compare results of the numerical modeling with the experiment, we present  the angular momentum per atom $L_z/N$ [see Fig. \ref{UbOmega} (a)], the number of  vortices $\nu_v$ [see Fig. \ref{UbOmega} (b)], and number of antivortices $\nu_a$ [see Fig. \ref{UbOmega} (c)] inside of the condensate annulus at the end of the cycle when the barrier is ramped down to zero. The relaxation time from the nonequilibrium turbulent state to a metastable state with the persistent current takes a considerable time (comparable with the three-second life-time of the annular vortices). During one second of the experimental cycle such a transition usually is not completed and the number of annular vortices fluctuates substantially when vortices move towards the condensate edge. To minimize the contribution from these random fluctuations, we have calculated $\nu_v$ [shown in Fig. \ref{UbOmega} (b)] and $\nu_a$ [shown in Fig. \ref{UbOmega} (c)]  by averaging the number of annular vortices and antivortices over the period of $10$ ms before the end  and $10$ ms after the end of the stirring process. It is interesting to note that the number of detected annular vortices and antivortices {\em decays} for very high values of $U_b$. This is because, as mentioned above, a high-amplitude stirring beam creates a weak link and therefore breaks the potential barrier for the external vortices. As a result, a vortex from the outside of the ring readily enters the central hole without crossing the condensate annulus.

The white dashed curve in Fig. \ref{UbOmega} (a) presents  an extrapolated isoline corresponding to unitary value of the angular momentum per particle $L_z/N=1$. The dotted white line gives the isoline with $L_z/N=0.02$, corresponding to the threshold for annular vortex excitation, when a vortex line is placed at the external TF boundary of the annulus. We note that our results are in a good agreement with the experimentally measured \cite{Wright2013} threshold for vortex excitation especially for $\Omega>\Omega_c\approx 16$ Hz. For the lower angular velocities the experimentally measured threshold value of the barrier height $U_b$ is well below the predictions of our 2D model, as well as of the 1D model suggested in Ref. \cite{Wright2013}.  However, as we observed in our numerical simulations, even relatively small deviation $\delta R$ of the stirring beam center trajectory ($\delta R\approx 2\mu$m, which is of order of the precision of stable beam axis positioning reported in Ref. \cite{PhysRevLett.84.806}) from the ideal circle of radius $R=22.6 \mu$m  can significantly decrease the threshold barrier height $U_b$ for a slowly rotating stirrer.

The problem of the persistent current formation in a stirred toroidal condensate deserves a further theoretical and experimental investigation. It is intstructive to generalize our 2D theoretical model to a realistic 3D geometry and to account for the stochastic thermal fluctuations. Also, it is of interest to investigate, both experimentally and theoretically, the relaxation process to the metastable persistent current when the condensate is cleansed from the annular vortices depending on the position of the stirring beam.

\section{Conclusions}
We have investigated a condensate stirred by a small, tunable, penetrable barrier moving azimuthally at a fixed angular velocity. Our theoretical analysis reveals the microscopic mechanism of the large-scale persistent current generation and demonstrate a small-scale complex vortex dynamics.
For a slow motion of the narrow stirring beam vortex-antivortex pairs appear near the center of the stirrer in the bulk of the condensate. The antivortex moves spirally to the external surface of the condensate and finally decays into elementary excitations, while the vortex becomes pinned in the central hole of the annulus, adding  one unit to the topological charge of the persistent current. An outer edge  surface mode develops and breaks into the vortices entering the condensate annulus when the barrier angular velocity is above some critical angular velocity, in accordance with results of Bogoliubov analysis and the predictions of the surface model. Performing numerical modeling for various combinations of potential barrier height and angular velocity, we determine the conditions for creation of vortices which appear to be in good agreement with the experimental results.


\end{document}